\documentclass[conference]{IEEEtran}

\usepackage{cite}
\usepackage{amsmath,amssymb,amsfonts}
\usepackage{algorithmic}
\usepackage{graphicx}
\usepackage{textcomp}
\usepackage{xcolor}
\def\BibTeX{{\rm B\kern-.05em{\sc i\kern-.025em b}\kern-.08em
    T\kern-.1667em\lower.7ex\hbox{E}\kern-.125emX}}
\newcommand{\Z}{\mathcal{Z}}
\newcommand{\E}{\mathcal{E}}

\begin{document}

\title{The Four Levels of Fixed-Points\\in Mean-Field Models
}
\author{\IEEEauthorblockN{Sarath Yasodharan \thanks{Supported by a fellowship grant from the Centre for Networked Intelligence (a Cisco CSR initiative) of the Indian
Institute of Science, Bangalore.}}
\IEEEauthorblockA{
sarath@iisc.ac.in}
\and
\IEEEauthorblockN{Rajesh Sundaresan}
\IEEEauthorblockA{
rajeshs@iisc.ac.in}
}

\author{\IEEEauthorblockN{Sarath Yasodharan\IEEEauthorrefmark{1} and
Rajesh Sundaresan\IEEEauthorrefmark{1}\IEEEauthorrefmark{2}}
\IEEEauthorblockA{\IEEEauthorrefmark{1}ECE Department and \IEEEauthorrefmark{2}Robert Bosch Centre for Cyber Physical Systems,\\
Indian Institute of Science, Bangalore 560012, India}
}

\maketitle

\begin{abstract}
The fixed-point analysis refers to the study of fixed-points that arise in the context of complex systems with many interacting entities. In this expository paper, we describe four levels of fixed-points  in mean-field interacting particle systems. These four levels are (i) the macroscopic observables of the system, (ii) the probability distribution over states of a particle at equilibrium, (iii) the time evolution of the probability distribution over states of a particle, and (iv) the probability distribution over trajectories. We then discuss relationships among the fixed-points at these four levels. Finally, we describe some issues that arise in the fixed-point analysis when the system possesses multiple fixed-points at the level of distribution over states, and how one goes beyond the fixed-point analysis to tackle such issues.
\end{abstract}


\section{Introduction}
The purpose of this paper is to highlight four levels of fixed-points in mean-field interacting particle systems, and to discuss  the relationships among them. No new result is established. However, the unified perspective and the insights from the connections between the fixed-points are worthy of exposition.

Interacting particle systems specify the microscopic interactions of a particle with the other particles in the system, i.e., nature of interactions, their frequencies, and how they influence the state of the particle.
This microscopic description then leads to a collective macroscopic outcome, for example, the fraction of particles in each state at a given time or at equilibrium. In interactions of the \emph{mean-field} type, the evolution of the state of a particle depends on the states of the other particles only through the empirical measure, i.e., the distribution of the particles across states. Such mean-field models arise in the study of wireless local area networks (WLANs)~\cite{bianchi-98}, randomised algorithms in load balancing networks~\cite{mitzhenmaker-00,bramson2010randomized}, processor sharing systems~\cite{vasantam2018mean}, study of epidemic spread in a community~\cite{kermack1927contribution}, etc.

In the study of such systems, one often looks for {\em fixed-points}. Let $\gamma$ be some property associated with a tagged particle. For concreteness, in a WLAN, $\gamma$ could be the collision probability when a particular tagged node (particle) accesses the medium of communication. Intuitively, if the system is in equilibrium, then all the nodes may see collisions with the same probability $\gamma$. This creates a mean-field. The study of the collision probability of the tagged node, now viewed as a {\em response} to this mean-field, may yield an expression $G(\gamma)$ for the tagged node's collision probability. Consistency then demands that the node's appropriate collision probability $\gamma^*$ is one where the response to the mean-field $G(\gamma^*)$ equals $\gamma^*$, i.e., $\gamma^*$ is a fixed-point of $G$ and is the property that all the nodes are likely to experience at equilibrium, see~\cite{kumar-etal-06}. This provides a heuristic method to study the behaviour of such complex interacting particle systems.

To get this analysis machinery going, one needs a tractable property $\gamma$ and an analysis that generates the response $G(\gamma)$ to the mean-field arising from $\gamma$. In the WLAN analysis, this property $\gamma$ was the collision probability~\cite{kumar-etal-06}. This intuition has helped us analyse many natural and engineered systems. It works some times, fails some times, and we now have a good idea when it works and when it fails.

In this paper, we consider four properties of increasing levels of detail in a general mean-field interacting particle system, and describe the associated fixed-points. These four levels are (i) the macroscopic observables of the system, (ii) the probability distribution over states of a particle at equilibrium, (iii) the time evolution of the probability distribution over states of a particle, and (iv) the probability distribution over trajectories. We then discuss conditions under which the fixed-points exist and explore the inter-connections.

{\em Organisation}: Section~\ref{section:model} describes our finite-state mean-field interacting particle system. Section~\ref{section:fixed-points} describes the four levels of fixed-points. Section~\ref{section:discussion} discusses the relationships among these fixed-points. Section~\ref{section:conclusion} discusses issues related to multiple fixed-points.

\section{A mean-field model}
\label{section:model}
In this section we describe a finite-state mean-field interacting particle system and discuss some of its properties.

\subsection{The model}
\label{subsection:model}
Let there be $N$ particles. Each particle has a state. The particle states evolves over time in a Markovian fashion. Let $X^{(N)}_n(t)$ denote the state of the $n$th particle at time $t$ and let $\Z$, a finite set, denote the state space. The macroscopic behaviour of the system at time $t$ is encoded in the empirical measure
\begin{align*}
\mu_N(t) = \frac{1}{N}\sum_{n=1}^N \delta_{X^{(N)}_n(t)};
\end{align*}
$\mu_N(t)(z)$ is the fraction of particles in state $z$ at time $t$. To describe the evolution of the states of the particles, we consider a directed and connected graph $(\Z, \E)$. This graph encodes the set of all allowed transitions of a particle; whenever $(z,z^\prime) \in \E$, a particle in state $z$ can move to state $z^\prime$. For each $(z,z^\prime) \in \E$, we are given a function $\lambda_{z,z^\prime}$ on the space of probability distributions on $\Z$ (which we denote by $M_1(\Z)$ and view as a subset of $\mathbb{R}^{|\Z|-1}$). A particle at state $z$ at time $t$ moves to state $z^\prime$ at rate $\lambda_{z,z^\prime}(\mu_N(t))$. That is, the evolution of the state of a particle depends on the states of the other particles only through the empirical measure of the states of  all the particles. This description of the state evolution of the particles gives rise to two continuous-time Markov processes: (i) the Markov process $\{(X^{(N)}_n(t), 1\leq n \leq N), t \geq 0\}$ that describes the joint evolution of the states of all the particles, and (ii) the more convenient  Markov process $\{\mu_N(t), t \geq 0\}$ that describes the evolution of the empirical measure of the system of particles, i.e., the empirical measure process.

\subsection{Two examples}
\subsubsection{Markovian model of a WLAN}
\label{section:mac}
$N$ wireless nodes (or particles) access a common wireless channel. Since multiple nodes can try to access the channel at a given time slot, an attempt to transmit a packet by a node can either result in a collision or in a successful transmission. A basic distributed Medium Access Control (MAC) protocol implemented in the nodes can be summarised as follows~\cite[Section~7.3]{kumar-etal-wireless}.  Each node has a state belonging to the state space $\Z = \{0,1,\ldots, K\}$. The states represent aggressiveness of packet transmission. In each time slot, a node in state $i$ \emph{independently} attempts a packet transmission with probability $c_i/N$. The $O(1/N)$ scaling of the attempt probabilities results in $O(1)$ attempt probability for the entire system. In the basic implementation, $c_i = c_{i-1}/2$; thus state $0$ is the most aggressive state and state $K$ is the least aggressive state.

Let us write down the collision probability of a tagged node in state $z_0$ when the empirical measure of the system is $\xi$, i.e., the number of nodes in state $z$ is $N\xi(z)$, $z \in \Z$. Since nodes attempt independently, the probability that no other node (other than the tagged node) attempts in a given slot is
\begin{align*}
\left(1 - \frac{c_{z_0}}{N}\right)&^{N\xi(z_0)-1}  \prod_{z \in \Z, z \neq z_0}\left(1 - \frac{c_z}{N}\right)^{N\xi(z)} \simeq \exp\{-\langle c, \xi \rangle\}
\end{align*}
for large $N$, where $\langle c, \xi \rangle = \sum_z (c_z/N)(N\xi(z)) = \sum_z c_z \xi(z)$ can be interpreted as the system attempt probability. A collision occurs when at least one other node makes a transmission; hence the collision probability is $\gamma = 1-\exp\{-\langle c, \xi \rangle\}$. Observe that the collision probability of a node depends on only $\xi$ rather than the individual states of the particles. The collision probability is an example of a macroscopic observable because one can observe the received power in the wireless channel and assess how many slots suffer collisions.

We now describe the evolution of a particle's state. Consider a tagged node in state $i$. If the node encounters a collision, it moves to a less aggressive state, i.e., to state $i+1$, and continues to attempt transmission of its head-of-the-queue packet with probability $c_{i+1}/N$.  On the other hand, if the node makes a successful packet transmission, it moves to state $0$ and follows the same protocol for the next packet in its queue. When the node is in state $K$, it continues to attempt with probability $c_{K}/N$ until a successful transmission when it moves from state $K$ to $0$; see Figure~\ref{fig:transitions}, but with an additional self loop at state $K$. Based on the approximation to the collision probability in the previous paragraph, the probability that the node moves from state $i$ to state $i+1$ is
$
(c_i/N) (1 - \exp\{-\langle c, \xi \rangle\})
$
and the probability that the node moves from state $i$ to state $0$ is
$
(c_i/N) \exp\{-\langle c, \xi \rangle\}
$. The continuous-time caricature for the above  WLAN is obtained by scaling the time-slot durations  to be $1/N$ in the above discrete-time description. The rate at which a node in state $i$ moves to state $i+1$ is

{\small
\begin{align*}
\lambda_{i,i+1}(\xi) = \frac{(c_i/N) (1 - \exp\{-\langle c, \xi \rangle\})}{1/N}  = c_i(1 - \exp\{-\langle c, \xi \rangle\})
\end{align*}
}%
and the rate at which a node in state $i$ moves to state $0$ is

{\small
\begin{align*}
\lambda_{i,0}(\xi) = \frac{(c_i/N)(\exp\{-\langle c, \xi \rangle\})}{1/N}  =  c_i \exp\{-\langle c, \xi \rangle\}.
\end{align*}
}%
With these transition rates, the continuous-time Markovian caricature is an example of the mean-field model described in Section~\ref{subsection:model}. The transition graph for this continuous-time model is shown in Figure~\ref{fig:transitions}.

\setlength{\unitlength}{1.7pt}
\begin{figure}
\centering
\begin{picture}(40,15)(0,0)
\thicklines
\put(-20,15){\circle{9}}
\put(0,15){\circle{9}}
\put(20,15){\circle{9}}
\put(56,15){\circle{9}}
\put(-21,13.5){$0$}
\put(-1,13.5){$1$}
\put(19,13.5){$2$}
\put(54,13.5){$K$}
\put(-15.5,15){\vector(1,0){10.5}}
\put(4.5,15){\vector(1,0){10.5}}
\put(24.6,15){\vector(1,0){6}}
\qbezier(0,10)(-10,5)(-20,10)
\qbezier(20,10)(-5,0)(-20,10)
\qbezier(56,10)(0,-10)(-20,10)
\put(-11,7.5){\vector(-1,0){0.1}}
\put(-3,5){\vector(-1,0){0.1}}
\put(9,-0.2){\vector(-1,0){0.1}}
\put(33,15){\circle*{1}}
\put(41,15){\circle*{1}}
\put(49,15){\circle*{1}}

\end{picture}
\caption{Allowed transitions in a continuous-time caricature of the WLAN.}
\label{fig:transitions}
\end{figure}
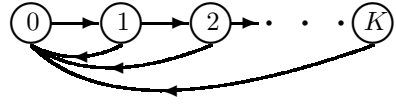

\subsubsection{The SIS Model of spread of an infectious disease}
Let there be $N$ individuals in a community. An infection spread in this community can be modelled using our mean-field system as follows. At any given time, each individual is in one of the two states: susceptible (S) or infected (I). Individuals in state S are healthy but can potentially catch the infection. Individuals in state I are infected. The individuals interact with each other and these result in the evolution of the states of the individuals as follows.  Whenever a susceptible individual comes in contact with an infected individual, the state changes from S to I. An individual in state I moves to state S after a random amount of time. Assuming \emph{uniform interaction} among the individuals in the community, the rate at which a susceptible individual becomes infected depends only on the fraction of infected individuals in the population. Therefore, this simple model of an epidemic spread is also of the mean-field type. See~\cite{kermack1927contribution} for an early study for such models.

\subsection{Law of large numbers}
Often in practice, properties that are relevant for the system (e.g., throughput in the WLAN example, equilibrium distribution in the SIS model) depend only on the macroscopic behaviour of the system and can be obtained from the empirical measure $\mu_N(t)$. Thus it is  important to study the empirical measure process $\mu_N$ in some detail. Towards this, we discuss a law of large numbers result for $\mu_N$~\cite{benaim-leboudec-08,bordenave-etal-12}.

Assume that the transition rates $\lambda_{z,z^\prime},\, \, (z,z^\prime) \in \E$, are Lipschitz continuous on $M_1(\Z)$. Assume that the initial conditions $\{\mu_N(0)\}_{N \geq 1}$ converge (in $M_1(\Z))$ to a probability distribution $\nu \in M_1(\mathcal{Z})$. Then for any fixed $T > 0$, the empirical measure process $\{\mu_N(t), 0 \leq t \leq T\}$ converges in probability to the solution to the ODE
\begin{align}
\dot{\mu}(t) = \Lambda_{\mu(t)}^* \mu(t), \, 0 \leq t \leq T, \, \mu(0) = \nu,
\label{eqn:MVE}
\end{align}
as $N \to \infty$; here, for any $\xi \in M_1(\mathcal{Z})$, $\Lambda_{\xi}$ denotes the $|\mathcal{Z}|\times |\mathcal{Z}|$ rate matrix  (i.e., $\Lambda_\xi(z,z^\prime)= \lambda_{z,z^\prime}(\xi)$ when $(z,z^\prime) \in \mathcal{E}$, $\Lambda_{\xi}(z,z^\prime) = 0$ when $(z,z^\prime) \notin \mathcal{E}$, and $\Lambda_{\xi}(z,z) = -\sum_{z^\prime \neq z} \lambda_{z,z^\prime}(\xi)$ for all $z \in \mathcal{Z}$) when the  empirical measure is $\xi$, $\Lambda^*_\xi$ denotes its transpose, $\mu(t)$ is the empirical measure at time $t$ viewed as a column vector, $\dot{\mu}(t)$ is its time derivative, and the convergence is in the space of $M_1(\Z)$-valued trajectories on $[0,T]$. The above ODE is called the McKean-Vlasov equation, and it evolves on the space of probability distributions on $\Z$. Owing to this convergence result, one can view the empirical measure process as a small random perturbation of the McKean-Vlasov equation and, for large $N$, the properties of the empirical measure process can be well approximated by the corresponding properties of the McKean-Vlasov equation.
\subsection{Propagation of chaos}
\label{section:decoupling}
We now discuss a \emph{decoupling approximation}   (also known as \emph{asymptotic independence} property) for the mean-field model~\cite{sznitman-91}. Again, we assume that the transition rates are Lipschitz continuous and that the initial conditions of  particles are independent and identically distributed as $\nu$\footnote{As a consequence,  $\mu_N(0) \rightarrow \nu$ in distribution as  $N \to \infty$.}. Consider two tagged particles, $1$ and $2$. The decoupling approximation tells us that, for each $t > 0$, the random variable $(X^{(N)}_1(t), X^{(N)}_2(t)) \rightarrow (Y_1(t), Y_2(t))$ in distribution as $N \to \infty$, where both the marginals $Y_1(t)$ and $Y_2(t)$ are distributed according to the solution to the McKean-Vlasov equation~\eqref{eqn:MVE} at time $t$ with initial condition $\nu$, and $Y_1(t)$ and $Y_2(t)$ are independent. That is, the initial \emph{chaos} propagates over time\footnote{The initial condition of independence of the particle states at time $t = 0$ can be relaxed to exchangeability so long as $\mu_N(0) \to \nu$ in distribution as $N \to \infty$, where $\nu$ is a fixed initial condition.} in the large-$N$ limit even though particles interact with each other. This property also has an obvious extension to any finite number (but not depending on $N$) of tagged particles. A similar  decoupling approximation holds in the stationary regime~\cite{bordenave-etal-12}. These  decoupling approximations play a crucial role in framing the fixed-point equations; see Section~\ref{section:fixed-points}.

\section{The four levels of fixed-points}
\label{section:fixed-points}
We now describe fixed-point equations at four different levels that arise in the context of our mean-field model. These fixed-point equations are formed based on the convergence and decoupling approximation results discussed in the previous sections. Solutions to these fixed-point equations describe the limiting behaviour (as $N \to \infty$) of the mean-field system and they are good approximations of large but finite-$N$ systems in some cases but not always.

\subsection{Level-1: Macroscopic observables}
\label{section:level1}
We start with the fixed-point equations that arise at the level of macroscopic observables of the system. While an abstract description of a fixed-point equation for generic macroscopic observables is possible, for concreteness, we use the WLAN described in Section~\ref{section:mac} as a running example. Recall the MAC protocol (with slotted time) and the Markovian continuous-time caricature described in Section~\ref{section:mac}.

Consider a tagged node in a system of $N$ nodes and let $\gamma$ denote its collision probability (at stationarity). That is, $\gamma$ is the probability that a packet transmitted by this node undergoes a collision. By symmetry of the model, this collision probability can be taken to be the same for all nodes.

We now compute the total attempt probability of the system as a response to the mean-field induced by $\gamma$~\cite{kumar-etal-06}. If $\beta(\gamma)$ is the attempt probability of the system, by symmetry, the attempt probability of the tagged node is $\beta(\gamma)/N$. Observe that the time instants at which the state of the tagged node becomes $0$ are renewal instants. Consider a renewal cycle between two successive visits to state $0$. Since the collision probability is $\gamma$ and the attempt probability in state $i$ is $c_i/N$, the mean renewal cycle duration (in number of time slots) is

{\small
\begin{align*}
 \frac{1}{c_0/N} + \frac{\gamma}{c_1/N}  + \cdots + \frac{\gamma^K}{c_K/N} + \left(\sum_{n \geq K+1} \frac{\gamma^{n}}{c_K/N}\right).
\end{align*}
}%
Similarly, the mean number of slots used for a successful packet transmission during a renewal cycle is
\begin{equation*}
1 + \gamma( 1 + \gamma (1+\gamma( \cdots )) \cdots).
\end{equation*}
Noting that the tagged node makes exactly one successful transmission in a renewal cycle, by the renewal-reward theorem, the attempt probability of the tagged node is
$
\frac{\beta(\gamma)}{N} = \frac{1+\gamma + \gamma^2 + \cdots}{\frac{N}{c_0} + \frac{\gamma N}{c_1} + \cdots + \frac{\gamma^K N}{c_K} + \frac{\gamma^{K+1}}{1-\gamma}\cdot \frac{N}{c_K}},
$
leading to

{\small
\begin{align}
\label{eqn:beta}
\beta(\gamma)= \frac{1+\gamma + \gamma^2 + \cdots}{\frac{1}{c_0} + \frac{\gamma}{c_1} + \cdots + \frac{\gamma^K}{c_K} + \frac{\gamma^{K+1}}{1-\gamma}\cdot \frac{1}{c_K}}.
\end{align}
}%
Let us now see the response of the tagged node when the mean-field is such that every other node sees a collision probability $\gamma$.  Using the decoupling approximation (see Section~\ref{section:decoupling}), the collision probability is

{\small
\begin{equation*}
1 - (1-\beta(\gamma)/N)^{N-1} \simeq 1- \exp\{-\beta(\gamma)\}
\end{equation*}
}%
for large $N$. Defining   $G(\gamma) = 1- \exp\{-\beta(\gamma)\} $, we see that $G(\gamma)$ is the response collision probability seen by the tagged node starting from an assumption that the collision probability $\gamma$. Hence self consistency demands that
\begin{align}
\gamma^* = G(\gamma^*),
\label{eqn:fixed-point-gamma}
\end{align}
which is a fixed-point of $G$. It can be checked that~\eqref{eqn:fixed-point-gamma} has a unique solution $\gamma^* \in (0,1)$ when the parameters $c_k$ are decreasing in $k$~\cite{kumar-etal-06}. For WLANs with a few tens of nodes and with the standard exponential back-off, i.e., $c_i = c_{i-1}/2$, $i = 1,2,\ldots, K$, this approximation works well in practice and $\gamma^*$ is close to the collision probability observed in practice. Thus, the  fixed-point analysis  at the level of macroscopic observables provides a simple and elegant way to characterise the performance of the WLAN system at equilibrium.

For a generic macroscopic observable for the mean-field model in Section~\ref{subsection:model}, one can come up with a fixed-point equation involving this observable by using the decoupling approximation in a suitable manner.
\subsection{Level-2: Distribution over states}
We now describe a fixed-point equation in the space of (equilibrium) probability distributions on $\Z$.

At time $t$, suppose that the distribution of each particle's state is $\xi$. Further, suppose that the initial conditions of the particles are exchangeable and $\mu_N(0) \to \nu$ in distribution as $N \to \infty$, for some deterministic $\nu$. Invoking the decoupling approximation in the transient regime, the empirical measure of the system at time $t$ concentrates near $\xi$ for large $N$. As the particles evolve over time, even though the empirical measure changes it remains close to $\xi$. We may assume that the mean-field is $\xi$.  Consider a tagged particle. Let $m$ denote the equilibrium response map that maps a given mean-field to the equilibrium response of the tagged particle. That is, $m(\xi)$ denotes the equilibrium distribution of this tagged particle in response to the states of the particles  distributed as $\xi$. Since the tagged particle's transition rates are given by the rate matrix $\Lambda_{\xi}$ when the mean-field is $\xi$, $m(\xi)$  is the $m$ that solves
\begin{align}
\sum_{z^\prime: (z^\prime, z) \in \E}m_{z^\prime}&\lambda_{z^\prime,z}(\xi) = m_z \sum_{z^\prime: (z,z^\prime) \in \E}\lambda_{z,z^\prime}(\xi),\, \, z\in\Z.
\label{eqn:eq-responsemap}
\end{align}
In~\eqref{eqn:eq-responsemap}, the left-hand side (LHS) is the total probability flux into $z$ and the right-hand side (RHS) is the total probability flux out of $z$ for the tagged particle in response to the mean-field $\xi$. At equilibrium these are in balance, which is the condition~\eqref{eqn:eq-responsemap}, and $m(\xi)$ is the $m$ that solves~\eqref{eqn:eq-responsemap}.

Self consistency demands that the appropriate equilibrium distribution is the $\xi^*$ that solves
\begin{align*}
m(\xi^*) = \xi^*,
\end{align*}
a fixed-point equation on the distribution of states of the particle for the equilibrium response map $m$. Equivalently,
\begin{align}
(\Lambda_{\xi^*})^*\xi^* = 0.
\label{eqn:fixed-point-xi}
\end{align}
Existence of the fixed point would follow from Brouwer's fixed-point theorem if one can show that mapping $m$ is continuous.  If there is a unique fixed-point  $\xi^*$ for~\eqref{eqn:fixed-point-xi}, then one might expect the system to equilibrate at $\xi^*$. This is not always the case, as we shall explain in Section~\ref{section:conclusion}.
\subsection{Level-3: Flow of distribution over time}
\label{section:level3}
We now describe a fixed-point equation on the space of $M_1(\Z)$-valued trajectories for the dynamic response map that maps a given flow of the mean-field over time to the flow of the distribution over states of a particle over time under the given mean-field.

Let an $M_1(\Z)$-valued trajectory $\xi(\cdot)$ denote the time-evolution of the distribution of a tagged particle, i.e., the state of the particle at time $t$ is distributed as $\xi(t)$. Suppose that each particle's state evolution is prescribed by the same $\xi(\cdot)$. Then the empirical measure of the system of particles at time $t$ is approximately $\xi(t)$ and hence the time-trajectory of the mean-field is approximately $\xi(\cdot)$.  The decoupling approximation suggests that the particles evolve, approximately, in an iid fashion.  In response, the state   of the tagged particle evolves according to the rate matrix $\Lambda_{\xi(t)}$ at time $t$ under the mean-field $\xi(\cdot)$. This evolution produces a response probability distribution flow over time denoted $\mathcal{M}(\xi(\cdot))$ for the tagged particle in response to the mean-field being $\xi(\cdot)$, i.e., $\mathcal{M}$ is the dynamic response map.  Write $m(t) = \mathcal{M}(\xi(t))$. One can write an ODE for $m(t)$ based on the fluxes of probability distributions into and out of each state:
\begin{align}
\dot{m}(t)(z) & =  \hspace*{-.3cm} \sum_{z^\prime: (z^\prime, z) \in \E} \hspace*{-.3cm} m(t)(z^\prime) \lambda_{z^\prime,z}(\xi(t))
- m(t)(z) \hspace*{-.35cm} \sum_{z^\prime: (z,z^\prime) \in \E} \hspace*{-.3cm} \lambda_{z,z^\prime}(\xi(t)) \nonumber \\
& = (\Lambda_{\xi(t)}^* m(t))(z),\, \, z \in \Z, t \geq 0;
\label{eqn:mt-response}
\end{align}
here, the RHS is the difference between the total probability flux into and out of state  $z$ in response to the mean-field $\xi(t)$. Self-consistency then demands that
\begin{align}
\xi^*(\cdot) = \mathcal{M}(\xi^*(\cdot)),
\label{eqn:fixed-point-mut-path}
\end{align}
a fixed-point equation on the space of $M_1(\Z)$-valued trajectories for the response dynamics map $\mathcal{M}$. That is, we must have $m(t) = \xi^*(t) ~ \forall t \geq 0$. Substitution of this in~\eqref{eqn:mt-response} yields
\begin{align}
\dot{\xi}^*(t) = (\Lambda_{\xi^*(t)})^* \xi^*(t), \, \, t \geq 0,
\label{eqn:fixed-point-mut}
\end{align}
which is precisely the McKean-Vlasov equation~\eqref{eqn:MVE}.

If the transition rates are assumed to be Lipschitz continuous on $M_1(\Z)$, then it is easy to check that the mapping $\xi \mapsto \Lambda_{\xi}^* \xi$ is also Lipschitz continuous on $M_1(\Z)$. As a consequence, standard results from ODE theory imply that there exists a unique solution to the ODE~\eqref{eqn:fixed-point-mut} and hence there exists a unique fixed-point for~\eqref{eqn:fixed-point-mut-path}.
\subsection{Level-4: Distribution over trajectories}
We now describe a Level-4 fixed-point equation  over the space of probability distributions $D([0,\infty), \Z)$, which is the space of $\Z$-valued trajectories. Note that this is different from Level-3 in the following way. Level-3 refers to fixed-points in the space of  $M_1(\Z)$-valued trajectories over time whereas Level-4 refers to fixed-points in $M_1(D([0,\infty), \Z))$, the space of distributions over $D([0,\infty), \Z)$.

Let $Q \in M_1(D([0,\infty), \Z))$ denote the probability distribution of a typical particle on the space of $\Z$-valued trajectories. At each time $t$, the time-marginal of $Q$ is the distribution of the state of the typical particle at time $t$. More precisely, it is the \emph{push forward} $Q \circ \pi_t^{-1}$ of the measure $Q$ under the canonical coordinate projection map $\pi_t$; $Q \circ \pi_t^{-1}$ is a probability distribution on $\Z$. Assuming the  decoupling approximation,  the mean-field  approximately evolves as $\{Q \circ \pi_t^{-1}, t \geq 0\}$.

Now consider a tagged particle responding to the  mean-field $\{Q \circ \pi_t^{-1}, t \geq 0\}$. Let $\mathcal{T}(Q)$ denote the  law of the tagged particle's evolution over time. Note that $\mathcal{T}(Q) \in M_1(D([0,T), \Z))$, it is the response to the mean-field induced by $Q$, and it depends on $Q$ only through the flow $\{Q \circ \pi_t^{-1}, t \geq 0\}$. Since the flow is given by $ \xi(t) = \{Q \circ \pi_t^{-1}, t \geq 0\}$,  $\mathcal{T}(Q)$ is the law of the Markov process associated with the rates $\Lambda_{\xi(t)}$. Consistency demands that
\begin{align}
Q^* = \mathcal{T}(Q^*).
\label{eqn:fixed-point-martingale-problem}
\end{align}

In probability theory, the problem of existence and uniqueness of Markov processes given a  generator (in our case, the transition rates $\Lambda_{\xi(t)}$), or an extended generator, is known as the \emph{martingale problem}~\cite{stroock-varadhan}. In our case, the generator itself is flexible, is determined by $Q$, and we are looking for a special solution that fixes the map  $Q \mapsto  \mathcal{T}(Q)$. This is called a \emph{nonlinear martingale problem}~\cite{graham-00}. Let $Q^*$ be a fixed point. The analogue of the  Kolmogorov forward equation that governs the evolution of the time-marginals of $Q^*$ is a nonlinear ODE~\eqref{eqn:MVE}, the McKean-Vlasov equation. It turns out that there exists a unique fixed-point for~\eqref{eqn:fixed-point-martingale-problem} if the transition rates are uniformly Lipschitz continuous on $M_1(\Z)$~\cite{graham-00}.
\section{Connections and Discussion}
\label{section:discussion}
We now describe some connections between solutions to fixed-point equations at the four levels. We also discuss questions related to the goodness of Level-2 fixed-points.
\subsection{Level-1 and Level-2}
\label{section:level1-level2-connection}
To explain the connection between Level-1 and Level-2 fixed-points, let us consider the WLAN example. Let  $\xi^*$ be the unique Level-2 fixed-point.

We first compute the collision probability when the mean-field is at $\xi^*$. Consider a tagged node. Let $\gamma$ denote the collision probability of this node when the macroscopic behaviour of the system is $\xi^*$; this collision probability will be the same for all the nodes. Consider a tagged node in state $z_0$. As explained in Section~\ref{subsection:model}, the collision probability of the tagged node is
$\gamma \simeq  1 - \exp\{-\langle c, \xi^* \rangle\}$ for large $N$. 

On the other hand, the Level-1 fixed-point analysis already gives us a collision probability $\gamma^*$, which is a fixed-point of~\eqref{eqn:fixed-point-gamma}. We thus expect that the two are equal, i.e., $\gamma^* = 1-\exp\{-\langle c, \xi^* \rangle\}$. This can be seen  as follows. On one hand, as explained in Section~\ref{section:level1}, a renewal argument tells us that $\beta(\gamma^*)/N$ is the attempt probability of a tagged node. On the other hand, using the decoupling approximation, assuming that each node at state $z$ independently attempts with probability $c_z/N$, we see that the attempt probability of a tagged node is  $\langle c, \xi^* \rangle/N$ when the mean-field is $\xi^*$. Thus, we have $\langle c, \xi^* \rangle=\beta(\gamma^*)$, and hence $\gamma^* = 1-\exp\{-\langle c, \xi^* \rangle\}$. This identity $\langle c, \xi^* \rangle=\beta(\gamma^*) $ can indeed be checked using the expression of $\beta$ given in~\eqref{eqn:beta} and the fact that $\xi^*$ satisfies $(\Lambda_{\xi^*})^*\xi^* = 0$. Therefore, the Level-2 fixed-point, which is a distribution over the states of a particle, explains the Level-1 fixed-point which is a macroscopic observable.

Such connections between macroscopic observables and distribution over states of the particles appear in thermodynamics. The relationship between the macroscopic variables of collision probability and attempt probability and their connection to the Level-2 fixed-point in WLAN is analogous to the relationship between macroscopic variables like temperature, pressure, and volume of an ideal gas and their connection to the Maxwell-Boltzmann distribution that describes the distribution of the velocities of the gas molecules. The passage from statistical mechanics to thermodynamics is made rigorous by large deviation theory; minimisations of rate functions in large deviation theory are mapped to the variational principles of free energy minimisation in thermodynamics~\cite{ellis2006entropy}.
\subsection{Level-2 and Level-3}
We now describe the connection between the Level-2 and Level-3 fixed-points. Suppose that $\xi^*$ is a Level-2 fixed-point of~\eqref{eqn:fixed-point-xi}, i.e., $(\Lambda_{\xi^*})^*\xi^* = 0$. The RHS of the Level-3 fixed-point equation~\eqref{eqn:fixed-point-mut} implies that when we start the McKean-Vlasov equation at $\xi^*$, it stays at $\xi^*$ for all times. That is, $\xi(\cdot) \equiv \xi^*$ is a \emph{stationary} Level-3 fixed-point. Conversely, any stationary Level-3 fixed-point of~\eqref{eqn:fixed-point-mut} is a Level-2 fixed-point of~\eqref{eqn:fixed-point-xi}, since we must necessarily have $(\Lambda_\xi)^* \xi = 0$. Thus, Level-2 fixed-points are precisely those points in $M_1(\Z)$ that are stationary Level-3 fixed-points.

Further, suppose that there is the unique stationary Level-3 fixed-point and that all its non-stationary Level-3 fixed-points (which are trajectories) converge to $\xi^*$ as time becomes large. Then there is a unique Level-2 fixed-point and it can be obtained from any Level-3 fixed-point by taking the large-time limit. In this special case, the macroscopic behaviour of the $N$-particle noisy system (which is a noisy version of the Level-2 fixed-point) in the large-$N$ limit \emph{concentrates} on the stationary Level-3 fixed-point~\cite[Theorem~2.3]{bordenave-etal-12}.
\subsection{Level-3 and Level-4}
Suppose that $Q$ is a Level-4 fixed-point of~\eqref{eqn:fixed-point-martingale-problem}. (We omit the superscript * to reduce clutter). Then the time-marginal of $Q$ at time $t$, $Q(t) = Q\circ \pi_t^{-1}$, is the distribution of a tagged particle at time $t$ (which is also approximately the mean-field at time $t$ when we have large number of particles). This distribution over states changes over time  as  particles make transitions. As we did in Section~\ref{section:level3}, the infinitesimal change in the distribution at time $t$ can be computed by the difference between the total probability flux entering and exiting various states. For state $z$, this difference is $(\Lambda_{Q(t)}^*Q(t))(z)$. On the other hand, this infinitesimal change is precisely $\dot{Q}(t)(z)$, the $z$-component of the time-derivative of $Q(t)$. Writing this condition as a vector condition for all $z$, we get
\begin{align}
\dot{Q}(t) = \Lambda_{Q(t)}^*Q(t).
\end{align}
Thus $Q(\cdot)$ is a Level-3 fixed-point. Hence, time-marginals of Level-4 fixed-points give rise to Level-3 fixed-points.

Conversely, consider a Level-3 fixed-point $\{Q(t), t\geq 0\}$. Clearly, $Q \in \mathcal{T}(\{R: R \circ \pi_t^{-1} = Q(t), t\geq 0\})$, but the set $\mathcal{T}(\{R: R \circ \pi_t^{-1} = Q(t), t\geq 0\})$ can possibly contain other probability distributions on trajectories. However, it turns out that when the  transition rates $\lambda_{z,z^\prime}$ are Lipschitz continuous on $M_1(\Z)$, $\mathcal{T}(\{R: R \circ \pi_t^{-1} = Q(t), t\geq 0\}) = \{Q\}$~\cite{graham-00}. Thus, under the assumption that the transition rates are Lipschitz continuous on $M_1(\Z)$, there is a one-one correspondence between the Level-3 and Level-4 fixed-points.

\section{Discussion on multiple fixed-points}
\label{section:conclusion}
We now discuss some issues that cannot be answered by just Level-1 and Level-2 fixed-point analysis. Let us focus on Level-2 fixed-points. By virtue of the relationship between Level-1 and Level-2 fixed-points, the issues for Level-2 will raise similar issues at Level-1.

{\em Multiple fixed-points:}
Suppose that $\xi_1^*$ and $\xi_2^*$ are two Level-2 fixed-points, and assume that these are \emph{locally asymptotically stable}, i.e., Level-3 fixed-points starting in a small neighbourhood of $\xi_1^*$ (resp.~$\xi_2^*$) converge to $\xi_1^*$ (resp.~$\xi_2^*$) as time becomes large. Since both $\xi_1^*$ and $\xi_2^*$ are Level-2 fixed-points, they are equilibria for the macroscopic state of the system. Therefore, whenever the finite-$N$ system's empirical measure is close $\xi^*_1$ (resp.~$\xi_2^*$), it remains close to $\xi_1^*$ (resp.~$\xi_2^*$) for a long time. However, because of the randomness present in the finite-$N$ system, the behaviour of the system can eventually move to the equilibrium $\xi_2^*$ and remain there for a long time. As explained in Section~\ref{section:level1-level2-connection}, the macroscopic observables may either be close to those arising from $\xi_1^*$ or $\xi_2^*$, and which of these is appropriate is a question that the Level-2 fixed-point analysis cannot answer by itself.

{\em Limit cycles:} Even if there is a unique Level-2 fixed-point, another issue can arise in the presence of stable \emph{limit cycles}. A limit cycle is a set $\mathcal{L} \subset M_1(\Z)$ with the property that time-marginals of Level-3 fixed-points with initial conditions lying in $\mathcal{L}$ stays at $\mathcal{L}$ at all times. When there is a stable limit cycle, even if there is a unique fixed-point, the large time behaviour of the system may be in the neighbourhood of the stable limit cycle especially when the unique fixed-point is unstable. Again, the fixed-point analysis would predict $\xi^*$ as the equilibrium behaviour which would then be incorrect.

{\em A sufficient condition for large-time characterisation:}
One sufficient condition when a unique Level-2 fixed-point $\xi^*$ does indeed characterise the system behaviour, so that the Level-2 fixed-point analysis is indeed a good approach, is when the fixed-point is globally asymptotic stable for the McKean-Vlasov dynamics, see \cite{benaim-leboudec-08,bordenave-etal-12}. This involves an examination of the properties of the Level-3 fixed-points.

{\em Beyond the fixed-point analysis:}
In general, when there are multiple Level-2 fixed-points, limit cycles, and/or chaotic attractors, one has to go beyond the fixed-point analysis and do a finer analysis to understand macroscopic observables. In the above example, a sufficient condition on the Level-3 fixed-point worked. In general, one must study the \emph{large deviations} behaviour of the mean-field of the system. Here, one  quantifies probabilities of transitions between Level-2 fixed-points (which are of the form $\exp\{-\text{constant}\times N\}$ for appropriate constants). These can be used to define an \emph{entropy function} $s$ on $M_1(\Z)$ such that, under stationary, the probability that we find the mean-field of the system near $\xi$ is proportional to $\exp\{-Ns(\xi)\}$~\cite{borkar-sundaresan-12}. This entropy function $s$ is called the large deviations rate function for the system under stationarity. We also mention that the  existence of multiple Level-2 fixed-points  and limit cycles also results in slower convergence rate for a given finite-$N$ system to its stationarity~\cite{mypaper-1}. As a consequence, one has to be cautious in interpreting Level-1 and Level-2 fixed-points, particularly when these are not unique or when there are stable limit cycles or when there are chaotic attractors for the limiting dynamics.

{\em Short-term unfairness:}
In the context of WLANs, at the level of nodes, one can encounter  \emph{short-term unfairness}~\cite{ramaiyan-etal-08,bhattacharya-kumar-17}. There are parameter choices where the wireless channel gets clogged by a single node for a  long time which results in low throughput for the others. A transition then occurs, after which the situation is similar with another single node hogging the channel at the expense of the others, and so on. The connection between this phenomenon, chaoticity of an attractor, quasi-stationarity~\cite{pollett2008quasi} remains to be explored.


\bibliographystyle{IEEETran}
\bibliography{ncc2021_v2}
\end{document}